\documentclass[runningheads]{llncs}
\usepackage{graphicx}
\usepackage{amsfonts}
\usepackage{hyperref}
\hypersetup{
  pdfinfo={
  pdfproducer={},
  Title={},
  Subject={},
  Author={},
  }
}

\begin{document}
\title{QSMGAN: Improved Quantitative Susceptibility Mapping using 3D Generative Adversarial Networks with Increased Receptive Field}
\titlerunning{QSMGAN}
%
\author{Yicheng Chen\inst{1,2} 
\and
Angela Jakary\inst{2}
\and
Sivakami Avadiappan\inst{2}
\and
Christopher P. Hess\inst{2}
\and
Janine M. Lupo\inst{1,2}
}
\authorrunning{Y. Chen et al.}
%
\institute{UCSF/UC Berkeley Graduate Program in Bioengineering, University of California, San Francisco and Berkeley, California.\\
\email{\{yicheng.chen,janine.lupo\}@ucsf.edu} \\ 
\and
Department of Radiology and Biomedical Imaging, University of California San Francisco, San Francisco, California.\\
\email{\{angela.jakary,sivakami.avadiappan,christopher.hess\}@ucsf.edu}}
\maketitle              
\begin{abstract}
Quantitative susceptibility mapping (QSM) is a powerful MRI technique that has shown great potential in quantifying tissue susceptibility in numerous neurological disorders. However, the intrinsic ill-posed dipole inversion problem greatly affects the accuracy of the susceptibility map. We propose QSMGAN: a 3D deep convolutional neural network approach based on a 3D U-Net architecture with increased receptive field of the input phase compared to the output and further refined the network using the WGAN with gradient penalty training strategy. Our method generates accurate QSM maps from single orientation phase maps efficiently and performs significantly better than traditional non-learning-based dipole inversion algorithms. The generalization capability was verified by applying the algorithm to an unseen pathology--brain tumor patients with radiation-induced cerebral microbleeds.

\end{abstract}

\section{Introduction}
Quantitative susceptibility mapping (QSM) is a recent phase-based quantitative magnetic resonance imaging (MRI) technique that enables in vivo quantification of magnetic susceptibility, a tissue parameter that is altered in a variety of neurological disorders \cite{Liu2015Quantitative,Wang2017Clinical}. QSM has been shown to quantify changes in vascular injury, such as the formation of cerebral microbleeds (CMBs) over time, hemorrhage, and stroke \cite{Klohs2011Detection,Liu2012Cerebral}. Iron deposition in the deep gray matter due to aging or disease can also be investigated using QSM\cite{Li2014Differential}. In neurodegenerative diseases such as Parkinson’s disease\cite{He2015Region-specific}, Alzheimer’s disease\cite{Acosta-Cabronero2013In} and Huntington’s disease\cite{Bergen2016Quantitative}, QSM can quantify the paramagnetic iron deposition related to disease progression and potentially generate biomarkers for diagnosing and managing neurodegenerative disease patients.

Although QSM has been demonstrated to have great potential in both research studies and clinical practice, accurate and reproducible quantification of tissue susceptibility requires multiple steps of careful data processing including phase reconstruction, coil combination (for multi-channel coils)\cite{Hammond2007Development}, multi-echo phase combination (for multi-echo sequences)\cite{Eckstein2017Computationally}, background phase removal \cite{Li2015iHARPERELLA:,Liu2011novel,Sun2014Background} and phase-susceptibility dipole inversion \cite{Li2015method,Liu2009Calculation,Liu2011Morphology}, Among them, the dipole inversion step is considered the most difficult because it is intrinsically an ill-posed inverse problem\cite{Deistung2016Overview}. This is due to the representation of the relationship between magnetic field perturbation and susceptibility distribution as a convolution, which can be more efficiently calculated as a point-wise multiplication in frequency space except along the conical surface where zero values result in missing data or noise amplification when solving for the inverse. To overcome this issue, missing data can be recovered by acquiring at least three scans with different relative orientations of the volume-of-interest in the main magnetic field, as with Calculation Of Susceptibility through Multiple Orientation Sampling (COSMOS)\cite{Liu2009Calculation}. This requires a subject to change head orientation between repeated scans, which has several disadvantages that significantly limit its application in practice: 1) the scan time is prolonged since multiple repeated scans are required, increasing both the cost of QSM and the risk of motion artifact; 2) co-registration of the different orientation images are required, which both increases processing time and introduces errors due to misalignment; and 3) modern high-field head coils are usually configured to be very close to the subject’s head for higher sensitivity, limiting the ability to rotate one’s head and degrading the quality of the QSM calculation. As a result, this approach is usually impractical for patient studies despite its superior potential to alleviate the ill-posed dipole inversion. 

Over the past decade, many approaches have been developed to address the ill-posed inverse problem. Thresholded K-space Division (TKD) simply thresholds the dipole kernel to a predetermined non-zero value to avoid dividing by zero\cite{Shmueli2009Magnetic}. Morphology Enabled Dipole Inversion (MEDI) regularizes the ill-posed inversion problem by imposing edge preservation constraints derived from magnitude images\cite{Liu2011Morphology}. Compressed Sensing Compensated inversion (CSC) exploits the fact that missing k-space data satisfies the compressed sensing requirement and applies a sparse L1 norm to regularize the problem\cite{Wu2012Whole}. Quantitative Susceptibility Mapping by Inversion of a Perturbation Field Model (QSIP) approaches the problem by inversion of a perturbation model and makes use of a tissue/air susceptibility atlas\cite{Poynton2015Quantitative}. These traditional methods suffer from three major limitations: 1) they either suffer from significant streaking artifacts or require careful hyperparameter tuning; 2) can be iterative and therefore take minutes to hours to compute; 3) they result in vastly different susceptibility quantifications, limiting reproducibility and making it difficult to compare studies that use different algorithms. 

Recently, Deep Convolutional Neural Networks (DCNNs) have shown great potential in computer vision tasks such as image classification\cite{He2016Deep}, semantic segmentation\cite{Long2015Fully} and object detection\cite{Ren2015Faster}. Among various deep neural network architectures, U-Net\cite{Ronneberger2015U-net:} has become the most popular backbone for many medical image-related problems\cite{Gong2018Deep,Kleesiek2016Deep,Zbontar2018fastMRI:} due to its effectiveness and universality. Bollmann et al.\cite{Bollmann2019DeepQSM} and Yoon et al.\cite{Yoon2018Quantitative} adopted the U-Net structure and extended it to 3D to solve the dipole inversion problem of QSM by training the network to learn the inversion using patches of various sizes as the input. Since its inception in 2014, Generative Adversarial Networks (GANs)\cite{goodfellow2014generative} have been incorporated into CNNs to further improve performance of segmentation, classification, and especially contrast generation tasks\cite{Arjovsky2017Wasserstein,Hammernik2018Learning,Nie2018Medical,Radford2015Unsupervised,Yang2018DAGAN:,Zhu2017Unpaired} by combining a generator that is trained to generate more realistic and accurate images with a discriminator that is trained to distinguish the real from the generated images. This idea of adversarial learning has recently been extended to applications in medical imaging\cite{Mardani2019Deep,schlemper2018stochastic,seitzer2018adversarial,Zhu2018Lesion,Zhu2019How}. The goals of this study were to for the first time: 1) incorporate the physical principles of the dipole inversion model that describes the susceptibility-phase relationship into the training of a deep neural network to generate QSM images and 2) harness the power of adversarial learning in this new application. We achieved these aims by: 1) modifying the structure of the 3D U-Net proposed by Bollmann et al. and Yoon et al.\cite{Bollmann2019DeepQSM,Yoon2018Quantitative} to incorporate an increased receptive field of the input phase image patches in conjunction with a cropping of resulting output in order to emulate the dipole physics within the structure of the model; and 2) by utilizing a GAN to regularize the model training process and further improve the accuracy of QSM dipole inversion.

\section{Materials and Methods}
\subsection{Theory of QSM dipole inversion and GANs}
Assuming that the susceptibility-induced magnetization is regarded as a magnetic dipole and the orientation of the main magnetic field $B_0$ is defined as the z-axis in the imaging Cartesian coordinate, the magnetic field perturbation and susceptibility distribution is related by a convolution, which can be efficiently calculated by a point-wise multiplication in frequency space\cite{Liu2015Quantitative}.
\begin{equation} \label{eq:1}
\Delta B_z(\mathbf{k}) = B_0(\frac{1}{3}-\frac{k_z^2}{|\mathbf{k}|^2})\chi(\mathbf{k})
\end{equation}

Where $\Delta B_z$ is the local field perturbation, $B_0$ is the main magnetic field, $\chi$ represents the tissue susceptibility, $\mathbf{k}$ is the frequency space vector and $k_z$ is the z-component. In practice, we measure $\Delta B_z$ by phase variation and solve the inverse problem for the susceptibility distribution $\chi$. However, notice that when $\frac{k_z^2}{|\mathbf{k}|^2}) \approx 1/3$, the bracket term on the right-hand side becomes close to zero, which causes missing measurements or noise amplification when solving the inverse problem, making it ill-posed. 

Assume $y(\mathbf{d}) = \Delta B_z(\mathbf{d})$ is the acquired tissue phase of the subject and x(d) is the susceptibility map of the subject we want to solve in the ill-posed phase-susceptibility dipole inversion problem, and function f represents the relationship between them, then we can simplify equation (\ref{eq:1}) with:

\begin{equation} \label{eq:2}
y=f(x)
\end{equation}

To solve the dipole inversion problem, we are finding a function $h$ that gives:

\begin{equation} \label{eq:3}
\tilde{x} = h(y)
\end{equation}

where $\tilde{x}$ is an estimate of the true susceptibility map $x$. The idea of GANs is to define a game between two competing components (networks): the discriminator (D) and the generator (G). G takes an input and generates a sample that D receives and tries to distinguish from a real sample. The goal of G is to “fool” D by generating more realistic samples. In this case, we use G as the function h:

\begin{equation} \label{eq:4}
\tilde{x}=G(y)
\end{equation}

The adversarial game between G and D is a minimax objective:

\begin{equation} \label{eq:5}
\min_G\max_D\mathbb{E}_{x\sim\mathbb{P}_q}[\log D(x)]+\mathbb{E}_{y\sim\mathbb{P}_t}[\log(1-D(G(y)))]
\end{equation}

where $\mathbb{P}_q$ is the distribution of true susceptibility maps and $\mathbb{P}_t$ is the distribution of tissue phases. To stabilize the training process, we adopt the method of Wasserstein GAN (WGAN)\cite{Arjovsky2017Wasserstein}, and the value function for WGAN is:

\begin{equation} \label{eq:6}
\min_G\max_{D\in \mathbf{D}}\mathbb{E}_{x\sim\mathbb{P}_q}[D(x)] - \mathbb{E}_{y\sim\mathbb{P}_t}[D(G(y))]
\end{equation}

where $\mathbf{D}$ is the set of 1-Lipschitz functions, which can be enforced by adding a gradient penalty term to the value function\cite{Gulrajani2017Improved}:

\begin{equation} \label{eq:7}
\lambda_{\mathrm{gp}}
\mathbb{E}_{y\sim\mathbb{P}_t}
(\Vert\nabla D(G(y))\Vert_2-1)^2
\end{equation}

where $\lambda_{\mathrm{gp}}$ is a parameter that controls the weight of the gradient penalty. Since the goal for G in this task is to recover/reconstruct QSM from a certain input tissue phase, we also included an L1 loss as content loss in the objective function of G:

\begin{equation} \label{eq:8}
\min_G\lambda_c\Vert x-G(y)\Vert_1 + \lambda_{\mathrm{adv}}L_{\mathrm{adv}}
\end{equation}

where $\lambda_{\mathrm{adv}}$ is the adversarial loss indicated in equation (\ref{eq:6}).

\subsection{QSMGAN framework}
\begin{figure}[ht]
  \centering
  \includegraphics[width=1.0\textwidth]{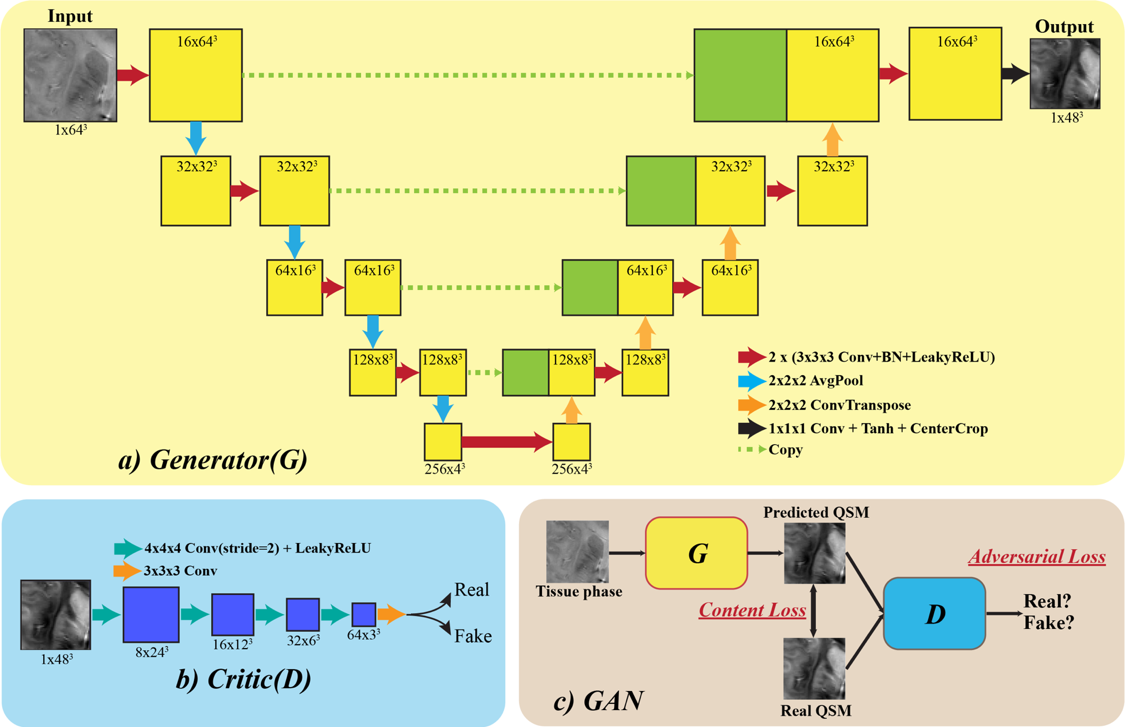}
  \caption{QSMGAN network architecture. a) The generator part of the GAN, which adopts a 3D U-Net with center cropping as a building block. b) The discriminator (“critic” in WGAN-GP) is constructed using 3D convolution with stride=2 to reduce image size. c) The overall GAN structure combines the generator and discriminator, where G is trained to generate more realistic and accurate QSM to fool D and D is trained to distinguish real and generated(fake) QSM.}
  \label{fig:fig1}
\end{figure}

We designed a 3D U-Net architecture similar to Bollmann et al. and Yoon et al. as the generator part of the QSMGAN framework as shown in Figure \ref{fig:fig1}. In each U-Net block, there are two 3x3x3 Conv3d-BatchNorm-LeakyReLU (negative slope of 0.2) layers, where Conv3d is the commonly used 3D convolution layer, the BatchNorm (batch normalization) accelerates and stabilizes the optimization and the LeakyReLU facilitates the training of the GAN. 3D average pooling was used to down-sample the image patch as proposed in the classic U-Net architecture, while 3D transpose convolution was applied to restore the resolution in the up-sampling path while incorporating the low frequency information back in the model. At the end of the generator, we applied a cropping layer to focus the training on only the center part of the patch. For the discriminator part of the QSMGAN, we designed a 3D patch-based convolutional neural network where each block of the network is composed of a 3D convolution (4x4x4 kernel size and stride 2) and a LeakyReLU (negative slope of 0.2). The four blocks in the network lower the input patch to 1/16 of the original size and the 3D convolution layer at the end converts the resulting patch to a binary output corresponding to the prediction of real and fake QSM patches. 

\subsection{Subjects and data acquisition}
Eight healthy volunteers (average age 28, M/F=3/5) were recruited for this study as the training and validation dataset for QSMGAN. All volunteers were scanned with a 3D multi-echo gradient-recalled sequence (4 echoes, TE= 6/9.5/13/16.5 ms, TR=50ms, FA=20, bandwidth=50kHz, 0.8mm isotropic resolution, FOV = 24 $\times$ 24 $\times$15cm) using a 32-channel phase-array coil on a 7T MRI scanner (GE Healthcare Technologies, Milwaukee, WI, USA). The sequence was repeated three times on each volunteer with different head orientations (normal position, tilted forward and tilted left) to acquire data for COSMOS reconstruction. GRAPPA-based parallel imaging \cite{Beatty2014Design} with an acceleration factor of 3 and 16 auto-calibration lines were also adopted to reduce the scan time of each orientation to about 17 minutes. 

To evaluate the generalization ability of our networks, we used a cohort of 12 patients with brain tumors who had developed CMBs years after being treated with radiation therapy. This type of vascular injury was an ideal pathology to test the generalizability of our network because they can both be extremely small in size and difficult to detect, and have very high susceptibility values compared to normal brain tissue due to deposits of hemosiderin. These patients were scanned using the same 7T QSM protocol as the healthy volunteers subjects except the slice thickness was 1.0mm. Only one orientation scan was performed on each patient. After the GRAPPA reconstruction, the image volumes were resampled to 0.8mm isotropic resolution to match the input of the deep learning models.

\subsection{QSM data processing and dataset preparation}
\begin{figure}[ht]
  \centering
  \includegraphics[width=1.0\textwidth]{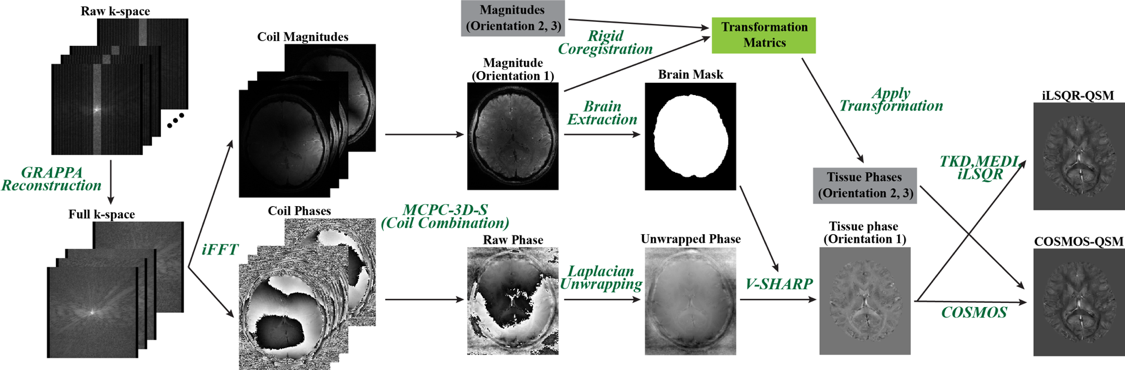}
  \caption{QSM data processing pipeline employed in this study. This figure shows processing of one scan orientation. Data from the other two orientations were processed similarly and introduced in the gray boxes in this figure to reconstruct the COSMOS-QSM. }
  \label{fig:fig2}
\end{figure}

The raw k-space data were retrieved from the scanner and processed on a Linux workstation using in-house software developed in Matlab 2015b (Mathworks Inc., Natick, MA, USA). The following processing steps (summarized in Figure \ref{fig:fig2}) were performed to obtain the tissue phase maps required for input to the QSMGAN and the calculation of the gold standard COSMOS-QSM which was used as the learning target of the QSMGAN: 1) GRAPPA reconstruction was applied to interpolate the missing k-space lines due to parallel imaging acceleration and channel-wise inverse Fourier transform was applied to obtain the coil magnitude and phase images; 2) coil images were combined to obtain robust echo magnitude and phase images using the MCPC-3D-S method\cite{Eckstein2017Computationally}; 3) raw phase was unwrapped using a Laplacian-based algorithm\cite{Li2011Quantitative}; 4) FSL BET\cite{Jenkinson2012FSL} was applied on magnitude images from all echoes to obtain a composite brain mask from the intersection of each individual echo mask; 5) V-SHARP\cite{Wu2012Whole} was used to remove the background field phase to get the tissue phase map; 6) images from different orientations were co-registered using magnitude images with FSL FLIRT\cite{Jenkinson2012FSL}; 7) the dipole field inversion was solved using the COSMOS algorithm\cite{Liu2009Calculation}. In addition, TKD\cite{Shmueli2009Magnetic}, MEDI\cite{Liu2011Morphology} and iLSQR\cite{Li2015method} QSM maps were also reconstructed from single orientation data for evaluation and comparison. A threshold of 0.15 was selected for the TKD algorithm, and $\lambda$=2000 was used in MEDI. The reconstructed single orientation tissue phase maps from the patient data were 1) used to compute iLSQR QSM and 2) fed into both the 3D U-Net and QSMGAN networks to generate COSMOS-like QSM. 

\subsection{Training and validation}
\begin{figure}[ht]
  \centering
  \includegraphics[width=0.7\textwidth]{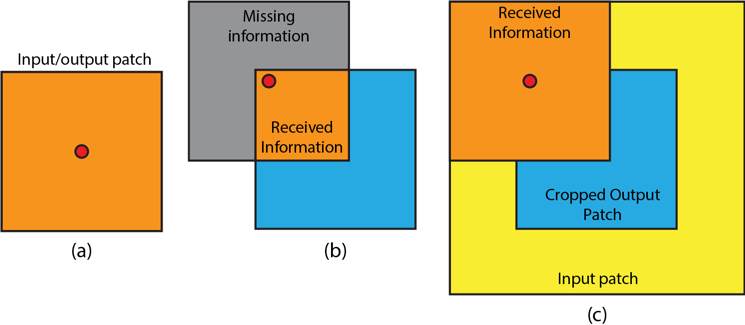}
  \caption{Demonstration of the relationship between receptive field and input/output patch size. a) Input patch size = output patch size, red dot represents voxels near the patch center. b) Input patch size = output patch size, voxels near the patch edge receive only information from the orange region. c) Input patch size > output patch size (with center cropping), voxels near the edge receive more information than in b).}
  \label{fig:fig3}
\end{figure}

The 8 healthy subjects were divided into 5 for training, 1 for validation, and 2 for testing. All three orientations were included in the dataset so the total number of scans in the training/validation/test set was 15/3/6. To build the training set, tissue phase and susceptibility patches were sampled by center coordinates with a gap of 8 voxels in all three spatial dimensions. Since background occupies most of the image volume, we sampled 90\% patches from inside the brain and only 10\% from the background to increase the efficiency of the training. For validation and testing, the input tissue phase volume was divided into non-overlapping patches according to the output patch size and the susceptibility map was reconstructed patch-wise by feeding the input tissue phase patch into the trained network. Figure \ref{fig:fig3} demonstrates the relationship between the receptive field and input/output patch size.

\begin{figure}[ht]
  \centering
  \includegraphics[width=0.7\textwidth]{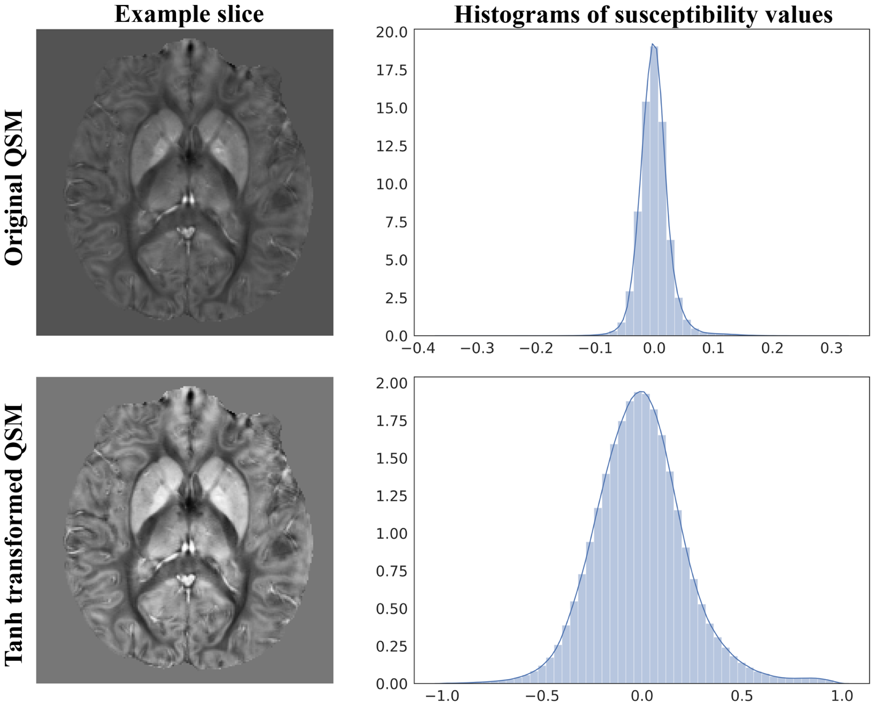}
  \caption{An axial slice of the original QSM (top left) and its histogram (top right) compared to the tanh transformed QSM (bottom left) and its histogram (bottom right). We can see that the tanh transform distributed the susceptibility values more evenly between -1.0 and +1.0, resulting in better contrast and value ranges for the network training. }
  \label{fig:fig4}
\end{figure}

To assist the neural network training, we multiplied the input phase by a scale factor of 100 and then transformed the output x by a scaled hyperbolic tangent operation to get the surrogate target $\dot{x}$:
$$\dot{x} = \mathrm{tanh}(10x)$$
This transform not only converts the range of the target susceptibility map to [-1, 1], which aids in the network training, but also results in a more Gaussian distributed histogram, helping the network learn values in different ranges (Figure \ref{fig:fig4}). 

As the baseline network, we first trained the U-Net based generator separately with the pairs of input and output patch sizes listed in Table \ref{tab:tab1}. To train the generator, an Adam optimizer with a learning rate of 1e-4 was used and betas were set to (0.5, 0.999). The network was trained for 40,000 iterations with a batch size of 16 that was lowered to 8 for larger input patch sizes. L1 loss was used as the loss function for the baseline network.

To train the QSMGAN, we again started with the baseline network and then: 1) fixed the generator G and trained D for 20,000 iterations to ensure that D was well trained, as suggested by \cite{Gulrajani2017Improved}; and 2) trained G and D together for 40,000 iterations. During each iteration, D (the critic) was updated 5 times with the gradient penalty $\lambda_{\mathrm{gp}}$=100. Adam optimizers were used for both G and D and the learning rate was lowered to 1e-5. To balance the content loss and adversarial loss, $\lambda_\mathrm{c}$ was set to 1 and $\lambda_{\mathrm{adv}}$ to 0.01.

\subsection{Evaluation metrics}
To evaluate the quality of the predicted QSM map reconstructed by the network ($\tilde{x}$), we calculated and compared the following metrics: 1) L1 error = $|x-\tilde{x}|_1$; 2) Peak Signal-to-Noise Ratio (PSNR) = $10\log_{10}\frac{R(x)}{\mathrm{MSE(x, \tilde{x})}}$, where $R()$ computes the voxel value range of the input image and MSE() computes the mean squared error between the reconstructed image and the target image; 3) Normalized Mean Squared Error (NMSE) = $\frac{\mathrm{MSE}(x, \tilde{x})}{|x|_2}$; 4) High-frequency error norm (HFEN); and 5) Structure similarity index (SSIM) as described in \cite{Langkammer2018Quantitative}. A Wilcoxon signed rank test was used to test for statistical significant differences in quality metrics between the optimized 3D U-Net and QSMGAN. 

Radiation-induced CMBs from each patient were segmented on reconstructed susceptibility weighted images (SWI) using in-house software\cite{Chen2018Toward,morrison2018user}. The resulting CMB masks were eroded by 1 voxel in all directions to remove the blooming artifact present on SWI and then applied to the iLSQR QSM, 3D U-Net and QSMGAN maps in order to quantify the median CMB susceptibility from the 3 different QSM images.  The number of CMBs were also counted for each patient using each of the 3 QSM maps by an experienced rater after blinded randomization of the images. A Kruskall-Wallis test was used to test for significant differences in median CMB susceptibility and CMB count among the 3 QSM methods and Bland-Altman Plots were used to visualize any discrepancies.

\section{Results}
\subsection{Baseline 3D U-Net}
\begin{figure}[ht]
  \centering
  \includegraphics[width=1.0\textwidth]{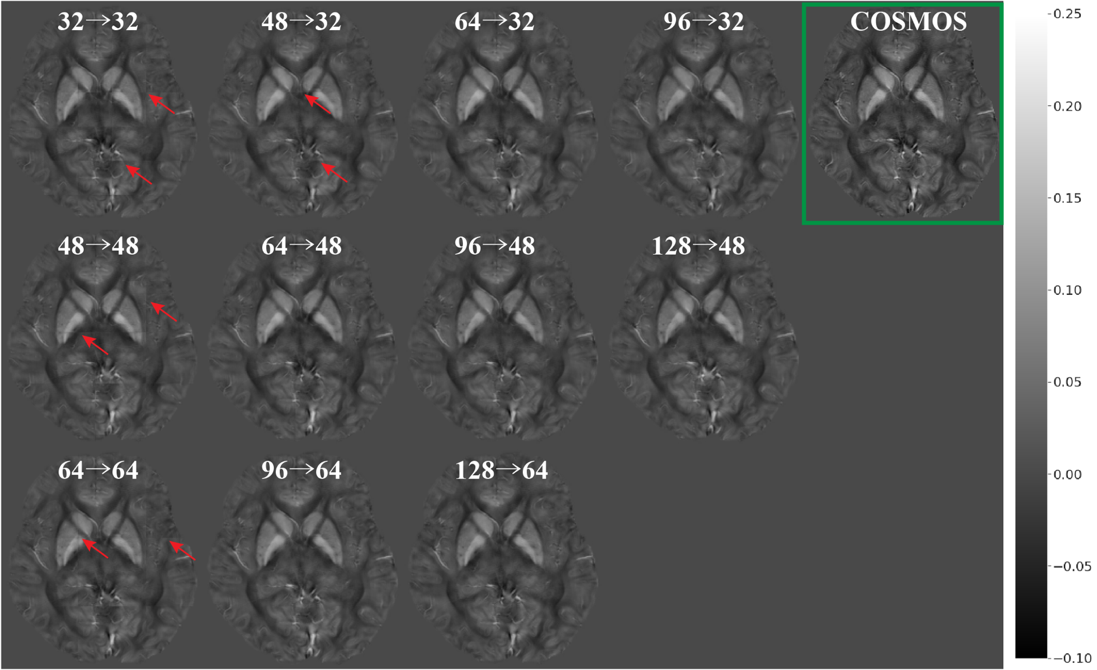}
  \caption{Comparison of reconstructed QSM using 3D U-Net with different input/output patch sizes (input$\rightarrow$output). The green box highlights the ground truth COSMOS QSM. Red arrows highlight the edge incontituity artifacts.}
  \label{fig:fig5}
\end{figure}

We experimented with combinations of three different input patch sizes ($32^3$, $48^3$, $64^3$) and 5 output patch sizes ($32^3$, $48^3$, $64^3$, $96^3$, $128^3$, with input$\rightarrow$output) for the baseline 3D U-Net. Figure \ref{fig:fig5} demonstrates the qualitative effects of different input-output size pairs (shown on axial slices) while Table \ref{tab:tab1} compares the quantitative metrics (L1, PSNR, NMSE) used to evaluate the quality of the resulting QSM maps. When the input patch size was the same as the output patch size, the inversion error increased towards the edge of the patch, resulting in visible discontinuities in a grid-like pattern in the reconstructed QSM map. The higher L1 error, lower PSNR and higher NMSE supports this phenomenon quantitatively. When we increased the input patch size and applied center cropping at the end of the U-Net as shown in Figure \ref{fig:fig3}, the patch edge artifact decreased and the metrics improved. Among the different combinations of patch sizes, the input patch size of $64^3$ and the output patch size of $48^3$ ($64\rightarrow48$) provided the best balance between sufficient accuracy of the U-Net dipole inversion and low computation burden/efficiency. Therefore, for the QSMGAN evaluation we used the $64\rightarrow48$ 3D U-Net as a basic building block.

\begin{table}
\centering
\begin{tabular}{c|ccc} 
\hline
3D U-Net Patch Size & L1 error (1e-3) & PSNR       & NMSE         \\ 
\hline
32$\rightarrow$32               & 1.490$\pm$0.184     & 42.25$\pm$1.01 & 0.302$\pm$0.056  \\
48$\rightarrow$32               & 1.403$\pm$0.204     & 43.07$\pm$1.22 & 0.252$\pm$0.063  \\
64$\rightarrow$32               & 1.316$\pm$0.230     & 43.39$\pm$1.37 & 0.237$\pm$0.072  \\
96$\rightarrow$32               & 1.319$\pm$0.216     & 43.38$\pm$1.32 & 0.237$\pm$0.068  \\ 
\hline
48$\rightarrow$48               & 1.424$\pm$0.195     & 42.58$\pm$1.13 & 0.281$\pm$0.061  \\
\textbf{64$\rightarrow$48} & \textbf{1.309$\pm$0.210} & \textbf{43.53$\pm$1.31} & \textbf{0.229$\pm$0.065}  \\
96$\rightarrow$48               & 1.310$\pm$0.212     & 43.37$\pm$1.28 & 0.237$\pm$0.067  \\
128$\rightarrow$48              & 1.311$\pm$0.215     & 43.40$\pm$1.31 & 0.236$\pm$0.068  \\ 
\hline
64$\rightarrow$64               & 1.389$\pm$0.211     & 42.87$\pm$1.21 & 0.264$\pm$0.063  \\
96$\rightarrow$64               & 1.316$\pm$0.207     & 43.46$\pm$1.28 & 0.233$\pm$0.066  \\
128$\rightarrow$64              & 1.322$\pm$0.211     & 43.32$\pm$1.27 & 0.240$\pm$0.067  \\
\hline
\end{tabular}
\caption{Test set performance of U-Net baseline with different input and output patch sizes.}
\label{tab:tab1}
\end{table}

\subsection{Effectiveness of QSMGAN}
Using the $64\rightarrow48$ 3D U-Net as the generator, the metric-wise benefit of using QSMGAN over the 3D U-Net is shown by the quantitative metrics listed in Table \ref{tab:tab2}. (p=0.03 for all metrics of 3D U-Net v.s. QSMGAN) Column 4 and 5 in Figure \ref{fig:fig6} and Figure \ref{fig:fig7} demonstrates the visual comparison of reconstructed QSM of 3D U-Net and QSMGAN, where the adversarial training further improved the quality of the reconstructed QSM map by reducing both residual blurring and the remaining edge discontinuity artifacts from the relatively smaller input patch size, providing a more accurate and detailed mapping of susceptibility compared to the 3D U-Net baseline.

\begin{table}
\centering
\begin{tabular}{cccccc} 
\hline
Methods      & L1 error(1e-3)       & PSNR                & NMSE                 & HFEN                & SSIM                  \\ 
\hline
TKD          & 2.826$\pm$0.178          & 38.82$\pm$1.69          & 0.496$\pm$0.076          & 99.84$\pm$4.86         & 0.806$\pm$0.023           \\
MEDI         & 2.909$\pm$0.194          & 41.24$\pm$1.71          & 0.539$\pm$0.059          & 100.99$\pm$5.02         & 0.912$\pm$0.027           \\
iLSQR        & 2.193$\pm$0.227          & 42.03$\pm$1.45          & 0.410$\pm$0.088          & 74.40$\pm$7.15          & 0.896$\pm$0.025           \\
U-Net 64$\rightarrow$48  & 1.309$\pm$0.210          & 43.53$\pm$1.31          & 0.229$\pm$0.065          & 48.45$\pm$8.30          & 0.944$\pm$0.018           \\
QSMGAN 64$\rightarrow$48 & \textbf{1.199$\pm$0.215} & \textbf{44.16$\pm$1.42} & \textbf{0.200$\pm$0.065} & \textbf{45.68$\pm$8.53} & \textbf{0.952$\pm$0.018}  \\
\hline
\end{tabular}
\caption{Test set performance of U-Net baseline, QSMGAN and non-learning-based algorithms.}
\label{tab:tab2}
\end{table}

\subsection{Comparison with non-learning-based methods}
\begin{figure}[ht]
  \centering
  \includegraphics[width=1.0\textwidth]{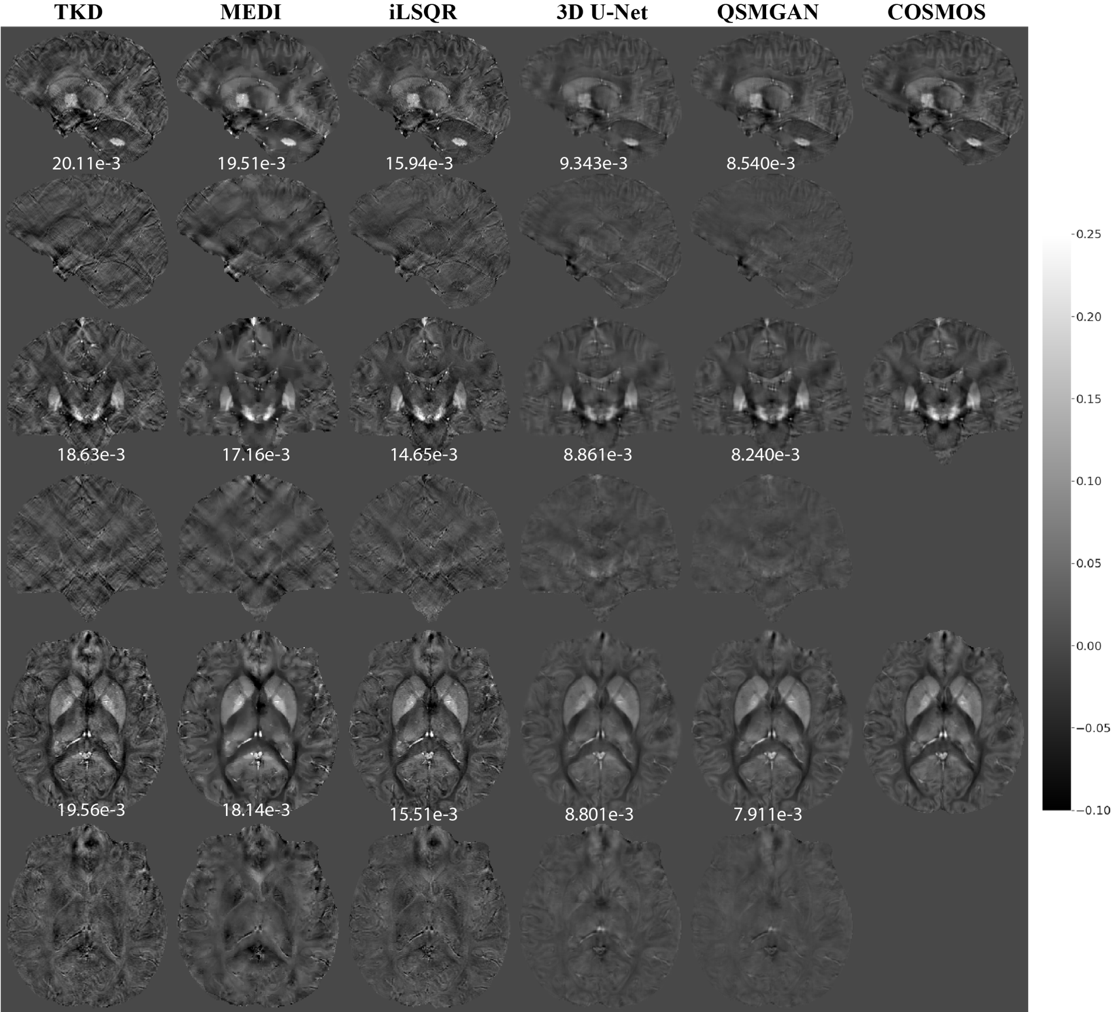}
  \caption{Comparison of QSM of test subject 1 reconstructed using non-learning-based dipole inversion algorithms (TKD, MEDI and iLSQR) and 3D UNet and QSMGAN. Row 1,2: sagittal view and error map. Row 3,4: coronal view and error map. Row 5,6: axial view and error map. Numbers at bottom of each slice show the L1 error relative to COSMOS-QSM of the slice.}
  \label{fig:fig6}
\end{figure}

\begin{figure}[ht]
  \centering
  \includegraphics[width=1.0\textwidth]{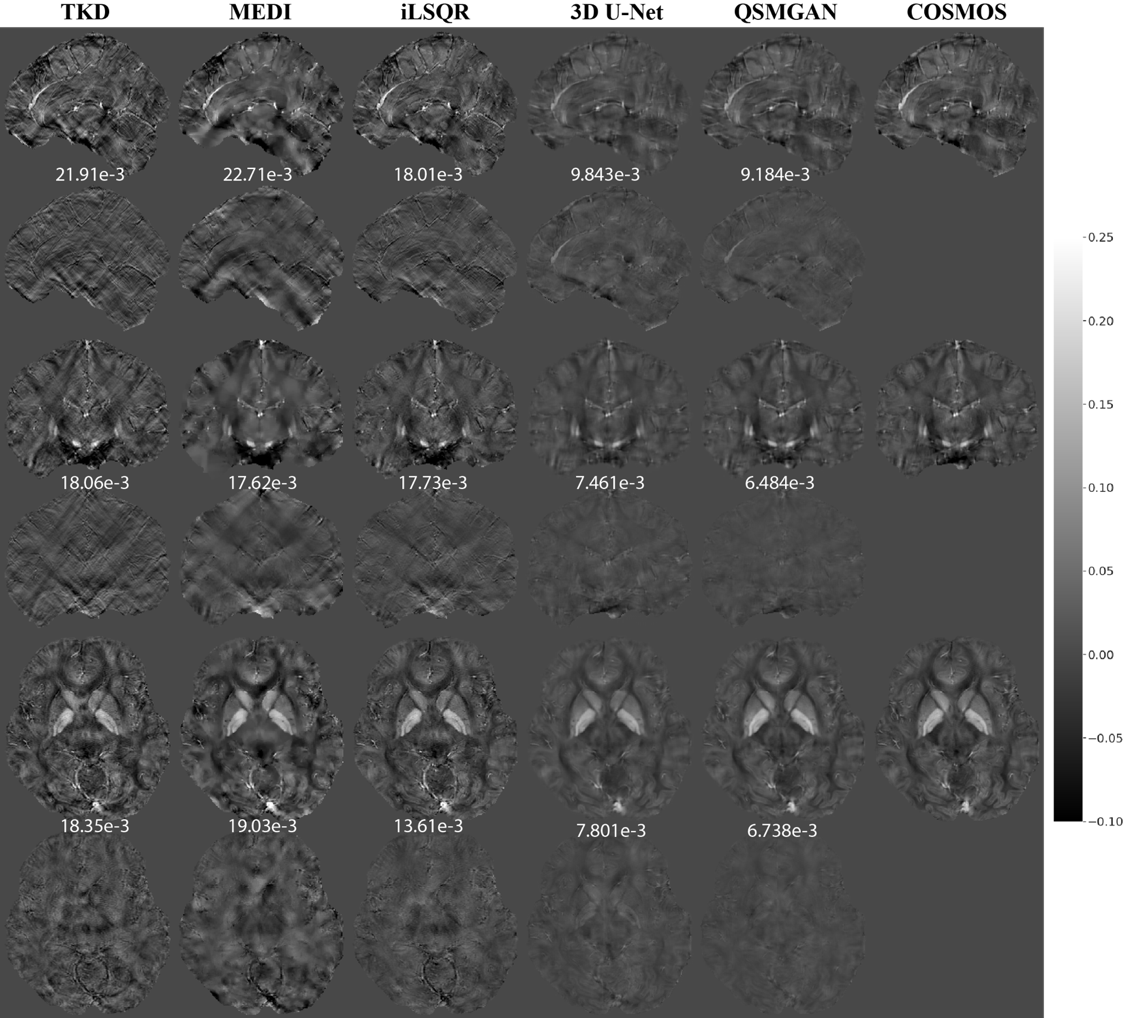}
  \caption{Comparison of QSM of test subject 2 reconstructed using non-learning-based dipole inversion algorithms (TKD, MEDI and iLSQR) and 3D UNet and QSMGAN. Row 1,2: sagittal view and error map. Row 3,4: coronal view and error map. Row 5,6: axial view and error map. Numbers at bottom of each slice show the L1 error relative to COSMOS-QSM of the slice.}
  \label{fig:fig7}
\end{figure}

Compared to 3 common ‘non-learning-based’ QSM dipole inversion algorithms (TKD, MEDI and iLSQR), our QSMGAN approach had 42-59\% reductions in NMSE and L1 error in the test datasets while increasing PSNR by 4-13\% as shown in Table \ref{tab:tab2}. Figure \ref{fig:fig6} and Figure \ref{fig:fig7} show examples of QSM slices from the two test subjects generated from our QSMGAN compared to non-learning-based algorithms. Although TKD had the lowest computational complexity, it also resulted in the most streaking artifacts. Despite its smooth appearance, MEDI was the least uniform with relatively high L1 error and inaccurate contrast of some fine structures such as vessels. It also required the longest computation time of all of the methods (about 2 hours on a regular desktop workstation). Although iLSQR QSM had lower L1 error than TKD and MEDI, it was visually noisier than all other methods. QSMGAN not only resulted in the best L1 error, PSNR, NMSE, HFEN, and SSIM, but achieved the most similar QSM map to COSMOS in only 2 seconds of reconstruction time per scan, the same order of time complexity as with the TKD method. 

\subsection{Application of networks in patients with radiation-induced CMBs}
\begin{figure}[ht]
  \centering
  \includegraphics[width=1.0\textwidth]{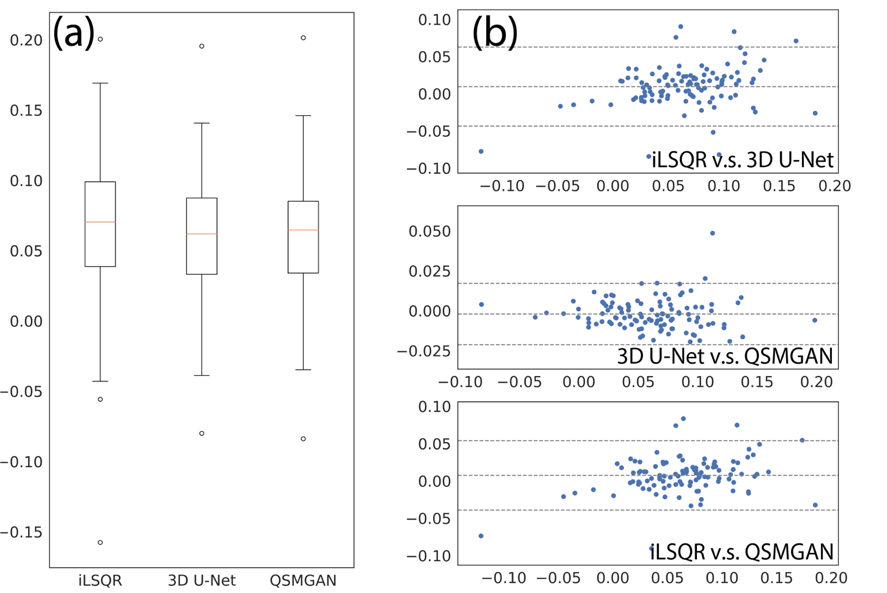}
  \caption{Comparison of median CMB susceptibilities measured from different QSM algorithms. a) box plot of median CMB susceptibilities. B) Bland-Altman plots of algorithm pairs.}
  \label{fig:fig8}
\end{figure}

\begin{figure}[ht]
  \centering
  \includegraphics[width=0.7\textwidth]{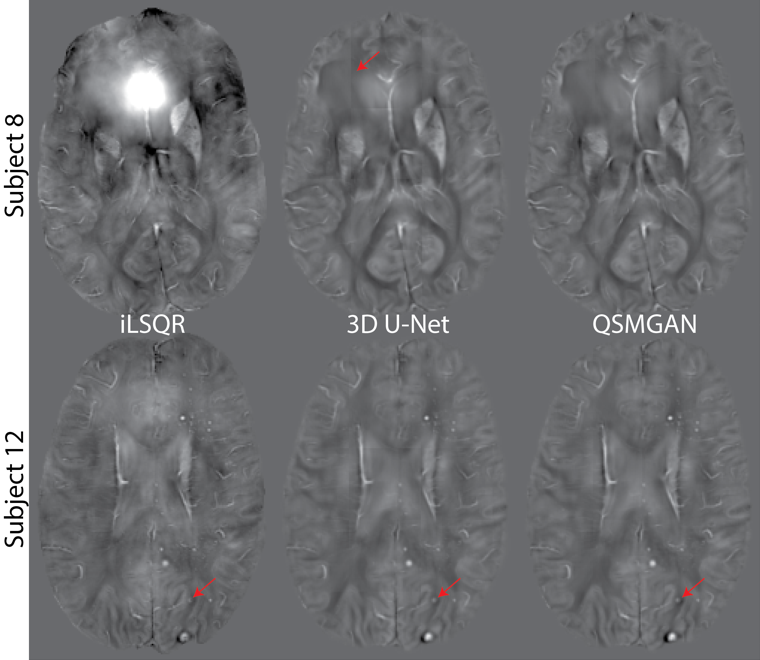}
  \caption{QSM of two patients with brain tumors who had developed cerebral microbleeds due to prior radiation therapy. Subject 8 suffered from poor brain extraction and background field removal that resulted in severe susceptibility artifacts in the iLSQR QSM image. Both 3D U-Net and QSMGAN sucessfully removed the artifact but QSMGAN generated higher quality maps with less edge discontinuity highlighted in the red arrow. Subject 12 had residual background phase that obscured the detection of a microbleed (denoted by the red arrow) that was correctly visualized on both the deep learning-based QSM maps.}
  \label{fig:fig9}
\end{figure}

To evaluate the generalization ability of our networks, we tested our network in a cohort of 12 patients with brain tumors treated with prior radiation therapy. The median susceptibility values for each CMB and total number of CMBs per patient based on iLSQR, 3D U-Net and QSMGAN were not significantly different among methods (Kruskal-Wallis test p=0.149 and p=0.936, respectively; see Figure \ref{fig:fig8}). This comparison demonstrates that the proposed QSMGAN could be well generalized to previously unseen pathology with extreme suscepbitility values. Figure \ref{fig:fig9} demonstrates the robustness of QSMGAN to artifacts from imperfect preprocessing steps such as skull stripping and background phase removal as well as its ability to generate more uniform susceptibility maps. Patient 8 (row 1) suffered from poor brain extraction and background field removal that resulted in severe susceptibility artifacts from the air-tissue interface in the sinuses in the iLSQR QSM image. Although 3D U-Net partially alleviated this problem, QSMGAN provided the most uniform and highest quality susceptibility map with the least amount of residual artifacts. Patient 12 (row 2) had residual background phase that obscured the detection of a microbleed (denoted by the red arrow) that was correctly visualized on both the deep learning-based QSM maps. 

\begin{table}
\centering
\begin{tabular}{cccc}
\hline
Subject & iLSQR & 3D U-Net & QSMGAN  \\ 
\hline
1       & 5     & 4        & 5       \\
2       & 14    & 3        & 3       \\
3       & 10    & 11       & 13      \\
4       & 12    & 9        & 7       \\
5       & 29    & 25       & 26      \\
6       & 18    & 15       & 15      \\
7       & 3     & 5        & 6       \\
8       & 4     & 5        & 4       \\
9       & 7     & 9        & 10      \\
10      & 10    & 11       & 13      \\
11      & 25    & 22       & 21      \\
12      & 25    & 23       & 22      \\
Total   & 158   & 142      & 145     \\
\hline
\end{tabular}
\caption{Cerebral microbleed counts from iLSQR 3D U-Net and QSMGAN. No significant difference was observed between the two methods.}
\label{tab:tab3}
\end{table}

\section{Discussion}
Although in theory the phase-susceptibility relationship in QSM is global, meaning the tissue phase is determined by the susceptibility of all locations in the imaging volume, we still adopted a patch-based deep learning approach similar to \cite{Yoon2018Quantitative} for several reasons. Since the network is 3D, the patch-based method can significantly reduce the computation complexity and memory requirement compared to whole-volume based approaches, especially when conducting high-resolution QSM. For example, if we needed to generate a full QSM volume with a 256x256x150 matrix size using the entire volume as an input to the 3D U-Net architecture, even the most advanced GPU with 32GB of graphics memory would not be able to fit a single training sample. The patch-based method also converts one single scan into hundreds of input images, even before data augmentation. Since COSMOS requires a relatively long scan time and is cumbersome to conduct, training a more generalizable deep convolutional network is beneficial when only a limited amount of data is available. Because the phase is mostly determined by nearby susceptibilities due to the properties of the susceptibility-phase convolutional kernel, the patch-based approach yields a good approximation of the dipole inversion. 

As Table \ref{tab:tab1} demonstrates, increasing the input patch size and applying center cropping at the end of the 3D U-Net significantly improved the quality of the reconstructed QSM maps. This can be intuitively described by Figure \ref{fig:fig3}, where when the input patch size equaled the output patch size, an output voxel near the center of the patch (Figure \ref{fig:fig3}a) could receive information from the entire patch. However, a voxel near the edge of the output patch (Figure \ref{fig:fig3}b) would only receive information from the orange region and a large portion of the phase information from the gray region would be missing, reducing the ability of the network to accurately solve for the susceptibility. When we increased the input patch size (Figure \ref{fig:fig3}c) and cropped the output patch such that only the center of the patch was considered a valid QSM prediction, voxels near the edge of the patch regained phase input information thereby increasing the accuracy of the quantified susceptibility values. 

Another observation from Table \ref{tab:tab1} is that the medium output patch size ($48^3$) achieved the best QSM reconstruction performance. The smaller patch size ($32^3$) performed worse because the output voxels received less information, introducing more error to the patch approximation of global convolution. Unexpectedly, the larger patch size ($64^3$) didn’t provide any extra benefit to the dipole inversion. This might be due to the fact that it introduced more variables into the computation process and increased the difficulty of training a good network for QSM reconstruction. In addition, for each output patch size, using excessively large input patches (such as $96\rightarrow 32$) did not further reduce the error but slightly downgraded the QSM quality. This might be due to increased information far from the output patch interfering with the dipole inversion. 

A disadvantage of using an excessively large input patch size is the dramatically increased computational complexity and GPU memory requirement. Note that the network is three-dimensional and the computational complexity and memory requirement of training the networks roughly increases with the input patch size by O($n^3$). The center cropping we applied to ensure a large enough receptive field, only exacerbated this problem, greatly reducing the efficiency of the prediction process. For example, if we increased the input patch size from $32^3$ to $64^3$, the training/prediction time and memory became 8x as long and only 1/8 of the computed patches were utilizied. Based on the observation that excessively large input patch sizes greatly increased the computational burden without improving the quality of the resulting QSM maps, we selected the $64\rightarrow48$ 3D U-Net as the base network to integrate with the GAN. 

The rationale for the GAN training, which included adding a discriminator or “critic”, was to guide the generator (or the 3D U-Net) to further refine its result so that it could not be distinguished from a real COSMOS QSM patch. Although it took a long time (48 hours) to train the QSMGAN, once the training was finished the discriminator was no longer needed. As a result, reconstruction or prediction of the QSM map for a new scan/subject from tissue phase only required one forward pass through the 3D U-Net for each input patch, thereby resulting in a computational complexity that is identical to the 3D U-Net baseline. 

Although the QSMGAN was trained only on healthy volunteer data, when applying the network to patient data, it successfully recovered the previously unseen pathology of cerebral microbleeds and assigned values similar to those obtained from iLSQR. This demonstrated that the networks avoided overfitting and managed to learn the underlying dipole convolution relationship between tissue phase and susceptibility sources. We also observed unforeseen robustness to imperfect preprocessing from QSMGAN. This was likely due to the fact that our QSMGAN was trained on carefully processed training data with little artifacts, so the generator would favor outputs with similar image quality and therefore tended to remove any abnormal suscepbility sources and remaining background phase components. Although our QSMGAN was trained on only brain images because QSM has been most widely utilized in the brain, the network can easily be trained using data from other organs of interest.

\section{Conclusions}
In this study, we implemented a 3D U-Net deep convolutional neural network approach to improve the dipole inversion problem in QSM reconstruction. To better approximate the global convolution property in the phase-susceptibility relationship through patch-based neural networks, we enlarged the input patch size and introduced center cropping to ensure an increased input receptive field for all neural network outputs. This cropping technique provided significantly lower edge discontinuity artifacts and higher accuracy. Including a generative adversarial network based on the WGAN-GP technique further improved the stability of training process, the image quality, and the accuracy of the susceptibility quantification. Compared to the other traditional non-learning dipole inversion algorithms such as TKD, MEDI and iLSQR, our proposed method could efficiently generate more accurate, COSMOS-like QSM maps from single-orientation, background-field-removed, tissue phase images. When tested on patients with radiation-induced CMBs, QSMGAN improved the robustness of the QSM reconstruction without sacrificing the sensitivity of CMB detection. Future directions include investigating the network’s ability to generalize to other scan parameters (such as TE, TR, and image resolution) and evaluating the performance of QSMGAN in patients with different pathologies and in other organs to ultimately improve patient care. 

\section{Acknowledgement}
This research is supported by NIH NICHD grant R01 NS099564.

%
%
%
\bibliographystyle{unsrt}
\bibliography{ref}
\end{document}